\documentclass[a4paper,10pt,aps,pre,twocolumn,superscriptaddress,footinbib,amsmath,amssymb]{revtex4-1}
\usepackage{hyperref}
\usepackage{graphicx,color}
\usepackage{textcomp}

\newcommand{\mrm}[1]{\mathrm{#1}}
\newcommand{\lsub}[0]{^{\vphantom{0}}}

\newcommand{\mean}[1]{\langle#1\rangle}

\newcommand{\Teff}[0]{T_\mathrm{eff}}
\newcommand{\Teffbar}[0]{\bar{T}_\mathrm{eff}}
\newcommand{\om}[0]{\omega}

\newcommand{\mh}[0]{m_\mrm{h}}
\newcommand{\omh}[0]{\omega_\mrm{h}}
\newcommand{\qh}[0]{q_\mrm{h}}
\newcommand{\dqh}[0]{\delta q_\mrm{h}}
\newcommand{\qhdot}[0]{\dot{q}_\mrm{h}}
\newcommand{\qhddot}[0]{\ddot{q}_\mrm{h}}
\newcommand{\Th}[0]{T_\mrm{h}}

\newcommand{\ml}[0]{m}
\newcommand{\ql}[0]{q}
\newcommand{\qldot}[0]{\dot{q}}
\newcommand{\qlddot}[0]{\ddot{q}}

\begin{document}

\title{
Driven Brownian particle as a paradigm for a nonequilibrium heat bath: Effective~temperature and cyclic work extraction
}

\author{R. Wulfert}
\thanks{These authors have contributed equally to this work.}
\affiliation{II. Institut f\"ur Theoretische Physik, Universit\"at Stuttgart, Pfaffenwaldring 57, 70550 Stuttgart, Germany}

\author{M. Oechsle}
\thanks{These authors have contributed equally to this work.}
\affiliation{II. Institut f\"ur Theoretische Physik, Universit\"at Stuttgart, Pfaffenwaldring 57, 70550 Stuttgart, Germany}

\author{T. Speck}
\affiliation{Institut f\"ur Physik, Johannes Gutenberg-Universit\"at Mainz,
  Staudingerweg 7-9, 55128 Mainz, Germany}

\author{U. Seifert}
\affiliation{II. Institut f\"ur Theoretische Physik, Universit\"at Stuttgart, Pfaffenwaldring 57, 70550 Stuttgart, Germany}

\begin{abstract}

We apply the concept of a frequency-dependent effective temperature based on the fluctuation-dissipation ratio to a driven Brownian particle in a nonequilibrium steady state.
Using this system as a thermostat for a weakly coupled harmonic oscillator, the oscillator thermalizes according to a canonical distribution at the respective effective temperature across the entire frequency spectrum.
By turning the oscillator from a passive ``thermometer'' into a heat engine, we realize the cyclic extraction of work from a \textit{single} thermal reservoir, which is feasible only due to its nonequilibrium nature.

\end{abstract}

\maketitle

\textit{Introduction.} - Cornerstone principles of equilibrium statistical mechanics, such as the equipartition of energy or the fluctuation-dissipation theorem (FDT) \cite{calle51}, are generally not directly applicable to ageing (\emph{e.g.}, glasses \cite{kurch05}) or driven systems (\emph{e.g.}, active particles~\cite{march13,bechi16,fodor16}).
Sometimes equilibrium relations can be reconciled by introducing an ``effective'' quantity that compensates for nonequilibrium deviations.
In this spirit, the FDT can be formally maintained by interpreting the fluctuation-dissipation ratio (FDR) as an effective temperature~\cite{hohen89,cugli97c,villa09,cugli11}.
However, a nonequilibrium FDR may depend on both time and the choice of observable,
which is fundamentally at odds with the properties of an equilibrium temperature.
This caveat has led to the common notion that the effective temperature acquires thermodynamical meaning only if these dependencies are not too pronounced or can be appropriated to separate length- and/or time scales~\cite{kurch05,cugli11,berth02,villa09,levis15,diete15}.

For this reason, the effective temperature concept has been so prolific in describing the nonequilibrium properties of glassy systems~\cite{cugli97c,cugli11}. 
While fast vibrational fluctuations remain equilibrated with the environment, the slow evolution of the out-of-equilibrium structure is characterized by a higher effective temperature, a scenario known as partial equilibration~\cite{cugli97c}. During the ageing process, this effective temperature slowly decreases until eventually the environmental temperature is reached on all time scales~\cite{kurch05,wang06}.
However, in complex fluids or biological matter this kind of dynamical time-scale separation is the exception rather than the rule. In this regime of mixed time scales, time-dependent FDRs have been studied, \emph{i.a.}, for active matter~\cite{chen07,loi08,loi11,palac10,gnoli14,levis15,turli16,fodor16}, sheared colloidal suspensions~\cite{berth02,speck09,lande12b} and single biomolecules~\cite{diete15}.
In a heuristic approach introduced in the glassy context~\cite{cugli97c,kurch05} and revisited for driven Brownian ~\cite{loi08,loi11,hayas07} and ageing~\cite{grade10} systems, the effective temperature is identified with the measurement by a thermometer. However, while in equilibrium any conceivable thermometer must read the same temperature, this is no longer true in the nonequilibrium case, where time scales matter. 
One way of dealing with time-dependent FDRs is to consider only time-integrated quantities. This yields an Einstein relation defining a long-term effective temperature $\Teffbar$, which can be measured, \emph{e.g.}, by a tracer particle with a long intrinsic time scale~\cite{hayas07,loi08,loi11,maggi14}.
The thermodynamical meaning of $\Teffbar$ on the macroscopic level emerges, \emph{e.g.}, in the description of sedimentation in active matter~\cite{palac10,szame14,ginot15}.
Alternatively, one can keep the effective temperature as a spectral quantity $\Teff(\om)$, in which case it governs the thermalization of a coupled subsystem having a unique eigenfrequency~\cite{cugli97c,kurch05}.

Recent studies have carried the issue of thermalization even further: What if a heat bath is itself driven into nonequilibrium?
In the classical domain, experimental realizations of nonequilibrium baths typically comprise active bacterial suspensions \cite{chen07,sokol09,leona10,maggi14,krish16}, while theory has focused on the coarse-grained stochastic dynamics and energetics of Brownian particles in nonequilibrium environments \cite{das02,kanaz12,maes14,szame14,steff16}. 
The bulk of the literature on nonequilibrium reservoirs, however, pertains to quantum heat baths that are engineered involving distinctly quantum-related effects, such as correlations~\cite{dille09}, coherence~\cite{scull01,scull03}, or squeezing~\cite{huang12,rossn14,manza16}.
The quantum Otto cycle~\cite{geva92}, typically featuring a harmonic oscillator as its working substance, has been established as a paradigmatic model system to study the ramifications of non-Boltzmannian reservoir statistics on the thermalization of a coupled subsystem.
Fueled by nonequilibrium reservoirs, its efficiency may exceed the thermodynamic Carnot limit~\cite{dille09,huang12,abah14,zhang14,rossn14,niede16}. 
Moreover, it allows for the continuous extraction of work from a single reservoir~\cite{scull01,scull03,manza16}.
The question whether such results hold true also for classical systems has been adressed only recently~\cite{niede16,krish16} and motivates the present study.

In the following, we first evaluate the FDR for a simple Brownian system driven into a NESS. We consider a time-dependent FDR and show that it provides an effective temperature in the sense that it quantifies the thermalization of a weakly coupled harmonic oscillator at arbitrary frequency, which in turn acts as a spectral thermometer.
Finally, putting the oscillator to work in a cyclic process, we demonstrate the continuous extraction of work from a single thermal reservoir. 
This intriguing behavior requires a non-monotonic decay of the effective temperature $\Teff(\om)$ with frequency $\om$. Identifying the energy exchange between oscillator and reservoir as heat allows us to quantify the efficiency of this process.

\begin{figure}
  \centering
  \includegraphics[width=.8\columnwidth]{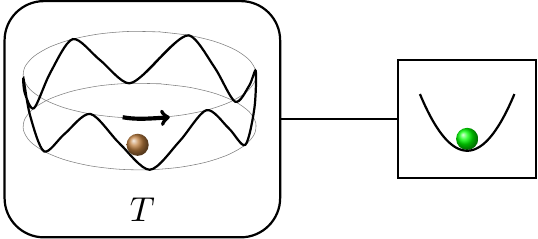}
  \caption{Scheme of the model. The coupling is realized by a linear velocity-dependent potential. The oscillator is used as a thermometer of the driven Brownian system.}
  \label{fig:model}
\end{figure}

\textit{Model.} -
Our model consists of two coupled subsystems, see Fig.~\ref{fig:model}.
The first part comprises a single underdamped Brownian particle in a bath at temperature $T$ moving along one dimension in a periodic potential 
\begin{align}
U(\ql)=U_0 \cos\left(\frac{2\pi}{L}\ql\right) ,
\end{align}
with $\ql$ denoting its position.
With periodic boundaries at positions $\{0,L\}$ the particle effectively travels on a ring.
A constant external force $f$ drives the system into a nonequilibrium steady state (NESS).

This system is in contact with a single particle in a harmonic potential
\begin{align}
V(\qh)=\frac{\omh^2}{2}\qh^2 ,
\end{align} 
which obeys Hamiltonian dynamics. 
In order for the system to act as a thermostat for the oscillator, we have to design a coupling that is weak in the thermodynamic sense, so that in equilibrium ($f=0$) the oscillator thermalizes to a canonical ensemble at bath temperature $T$.
In the case of a single particle subsystem, this is ensured if the coupling is bilinear and the coupling constant is sufficiently small~\cite{gelin09,cugli97c}.
We consider an interaction potential
\begin{align}
V_\mathrm{int}(\qldot,\qh)=\epsilon \qldot \qh ,
\end{align}
which depends on the velocity of the Brownian particle to circumvent difficulties of a position dependent coupling as encountered in reference \citep{hayas07} and justify using velocity-force FDR quantities.  

Including both the Langevin thermostat and the driving force $f$, the Brownian particle dynamics is given by
\begin{align}
\ml\qlddot=\epsilon\qhdot + U_0 \frac{2\pi}{L}\sin\left(\frac{2\pi}{L}\ql\right)+f-\gamma \qldot+\xi(t) .
\end{align} 
The stochastic force has zero mean and auto-correlations
\begin{align}
\left\langle \xi(t) \xi(t') \right\rangle =2 T \gamma \delta (t-t') ,
\end{align}
where $\gamma$ is the friction constant. 
Here and throughout, we set the Boltzmann constant $k_B=1$.
The equation of motion of the oscillator reads
\begin{align}
\mh\lsub \qhddot\lsub = - \epsilon\qldot - \omh^2 \qh\lsub.
\label{eq:dyn_osc}
\end{align}
The coupling thus shifts the average oscillator position to $\mean{\qh}=-\epsilon\mean{\qldot}/\omh^2$, which has to be taken into account when considering fluctuations $\dqh\equiv\qh-\mean{\qh}$.
All simulations are performed using a leap-frog algorithm with a time step of $\tau=0.0005$ and system parameters $T=1$, $m=\mh=1$, $\gamma=1$, $U_0=3$ and $L=4\pi/3$.
The coupling strength is set to $\epsilon=0.15$, which will later be shown to be small enough to ensure thermalization of the coupled subsystem.
\begin{figure}[b!]
\centering
 \includegraphics{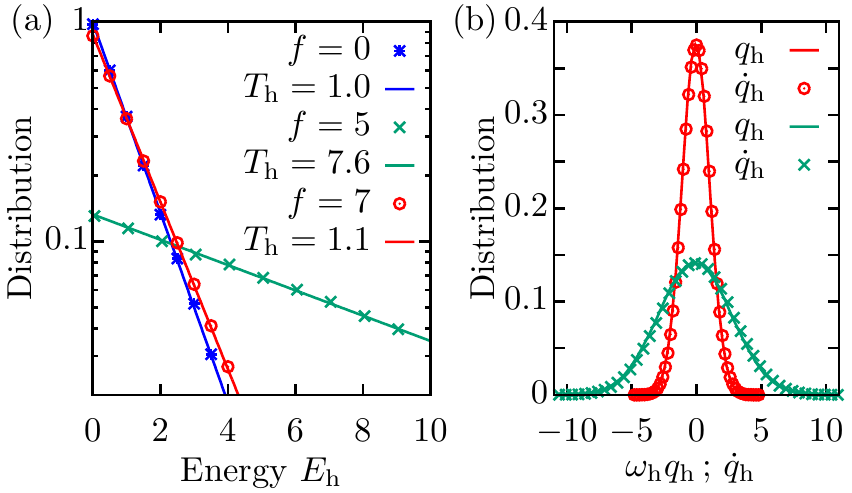}
\caption{(a)~Probability distribution of the internal energy $E_\mrm{h}$ of an oscillator with eigenfrequency $\omh=6.0$ sampled from simulations (symbols) of the equilibrium case ($f=0$) and two NESSs ($f= 5,7$). The data are fitted (lines) with respect to the oscillator temperature $\Th$ using Eq.~\eqref{eq:probdist_osc}.
(b)~Marginal distributions of rescaled oscillator position $\omh\qh$ (lines) and velocity $\qhdot$ (symbols) for driving forces $f=5$ (green; crosses) and $f=7$ (red; circles). The Gaussian distributions match in both cases, suggesting an equipartition principle holds with respect to quadratic terms in the internal oscillator energy.
}
\label{fig:hist}
\end{figure}

\textit{Effective Boltzmann distribution.} - We derive steady-state histograms of the oscillator having an internal energy $E_\mrm{h}(\qh,\qhdot) = \mh\lsub \omh^2 \qh^2/2 + \mh\lsub \qhdot^2/2$ and compare it to the corresponding Boltzmann distribution
\begin{align}
\rho(E_\mrm{h})=\frac{1}{\Th}\mathrm{exp}\left[ -\frac{E_\mrm{h}(\qh,\qhdot)}{\Th} \right], 
\label{eq:probdist_osc} 
\end{align}
where the oscillator temperature $\Th$ will serve as a fit parameter.
Apart from the equilibrium case ($f=0$), where for weak coupling Eq.~\eqref{eq:probdist_osc} becomes an exact relation with $\Th=T=1$ regardless of the oscillator frequency $\omh$, it is a priori unclear how the oscillator fluctuations will be affected by arbitrary driving $f$ as well as by the choice of $\omh$.
For now, we exemplarily choose $\omh=6.0$, before extending the analysis to arbitrary frequencies later.
Histograms fitted with Eq.~\eqref{eq:probdist_osc} are plotted in Fig.~\ref{fig:hist}(a).
The data have been averaged over 100 runs per $f$-value, recorded after the full system having reached a steady state.
In equilibrium ($f=0$) the oscillator thermalizes to a Boltzmann distribution at $\Th=1$, thus confirming the choice of $\epsilon$ to be sufficiently small.
Notably, the driven system also induces Boltzmannian statistics for the oscillator. 
The oscillator temperatures $\Th$, however, deviate significantly from $T=1$ and may now be specific to the oscillator frequency $\omh$.
For $f=5$, \emph{e.g.}, the distribution coincides with Eq.~\eqref{eq:probdist_osc} at $\Th\simeq7.62$.
Fig.~\ref{fig:hist}(b) shows the corresponding marginal distributions over $\omh\qh$ and $\qhdot$, which contribute quadratically to the oscillator energy $E_\mrm{h}$. 
Their virtually identical Gaussian forms suggest that the principle of energy equipartition is preserved under the nonequilibrium driving of the thermostat.
Also, Fig.~\ref{fig:hist}(b) shows that the coupling is weak enough, at least for the given $\omh=6.0$, not to induce a discernible shift $\mean{\qh}>0$ in the oscillator position.
In order to understand the underlying mechanisms that lead to this equilibrium-like thermalization and to clarify the role played by the oscillator frequency $\omh$, we will now turn to the fluctuations and linear-response properties of the Brownian system, which are captured by its steady-state FDR.

\begin{figure}[b!]
  \begin{center}
  \includegraphics[width=.9\columnwidth]{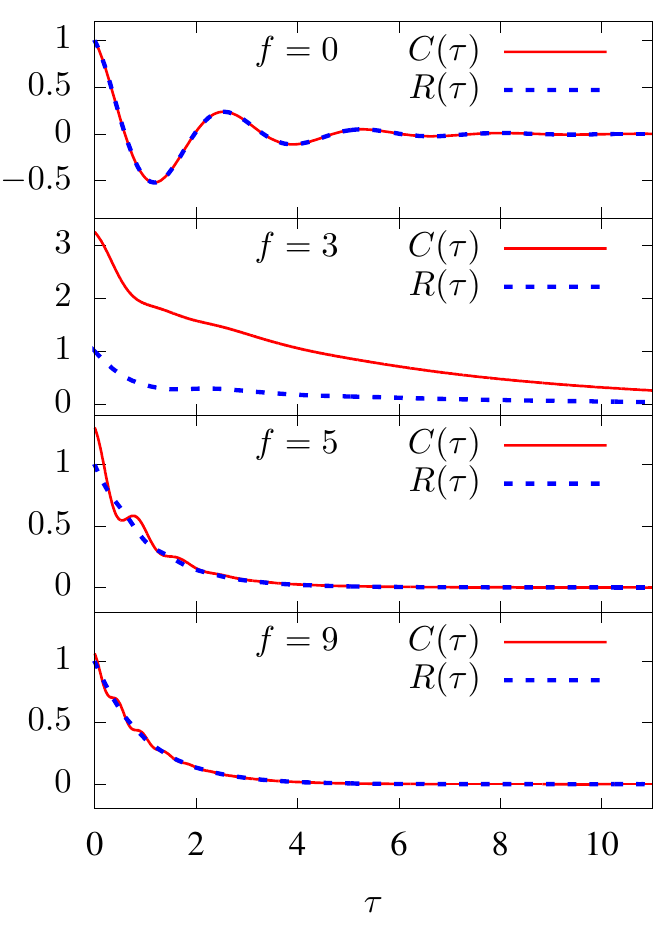}
  \end{center}
  \caption{Correlation and response functions for the equilibrium case ($f=0$) and three NESSs ($f=3,5,9$). In equilibrium at a bath temperature $T=1$, the FDT requires that $C(\tau)/R(\tau)=1$ for all $\tau$. Here, the oscillations in $C(\tau)$ and $R(\tau)$ reflect the underdamped motion in the potential wells. They are damped out by the driving for intermediate forces ($f=3$). In the running-state regime ($f=5,9$), oscillations reappear but are now caused by the impact of the periodic potential on the drifting particle.
  }
  \label{fig:corr_resp}
\end{figure}

\textit{Effective temperature from FDR.} - 
The frequency-dependent FDR defining the effective temperature~\cite{cugli97c} of the Brownian system reads
\begin{align}
\Teff(\omega) \equiv \frac{\tilde{C}(\omega)}{2\mathrm{Re} \bigl[\tilde{R}(\omega)\bigr]}
\label{eq:teff}
\end{align}
and contains the Fourier-transformed correlation and response functions $\tilde{C}(\omega)$ and $\tilde{R}(\omega)$ according to the velocity-force FDT.
Explicitly, we will consider the auto-correlation of velocity fluctuations
\begin{equation}
  C(t-t')=\left\langle \delta \qldot(t) \delta \qldot(t')\right\rangle
\end{equation}
and the associated linear response 
\begin{equation}
  R(t-t')= \left. \frac{\delta \left\langle \delta \qldot(t)\right\rangle}{\delta \zeta(t')} \right|_{\zeta=0}
\end{equation}
to a perturbative force protocol $\zeta(t')$.
In the processing of simulation data we will use the expression \mbox{$R(t-t')=\left\langle \delta \qldot(t) \xi(t')\right\rangle/2$}, which allows us to sample the response function from unperturbed steady-state trajectories~\cite{speck06}.
% shorten
In accordance with the FDR being an inherent property of the driven Brownian system, it is sampled without the oscillator attached, rendering it explicitly independent of any coupling or oscillator specifics.
For the weak coupling $\epsilon=0.15$, however, we have confirmed numerically that an attached oscillator does not impact the FDR.

For several driving forces $f$, Fig.~\ref{fig:corr_resp} shows correlation and response functions $C(\tau)$ and $R(\tau)$ in the time domain, which in the steady state depend on the interval $\tau\equiv t-t'$.
While in equilibrium the FDT \mbox{$\Teff=T=C(\tau)/R(\tau)$} is satisfied for all $\tau$,
the driving brings about time-dependent violations accompanied by pronounced qualitative changes ($f=3$).
For strong driving ($f=9$), the system exhibits equilibrium-like behavior with the FDR reapproaching the bath temperature $T=1$.
This quasi-equilibrium emerges as the driving dominates over any potential forces and is ultimately counteracted only by viscous drag.

\begin{figure}[t!]
\center
  \includegraphics{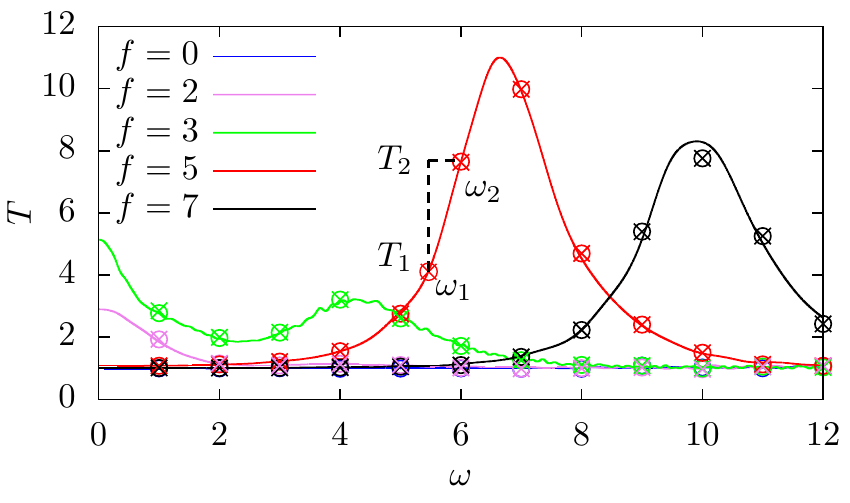}
  \caption{Frequency-dependent effective temperature $\Teff(\omega)$ (solid lines) for various driving forces $f$ in direct comparison with the kinetic and ``potential'' oscillator temperatures $\Th^\mrm{kin}$ (circles) and $\Th^\mrm{pot}$ (crosses). Standard deviations of the oscillator temperatures are always smaller than the symbol size. For $f=5$, the frequencies $\omega_1$ and $\omega_2$ and the respective values of the effective temperature $T_1$ and $T_2$ will be used in the cyclic protocol described at the end.}
  \label{fig:efftemp}
\end{figure}

\begin{figure}[b!]
\begin{center}
 \includegraphics{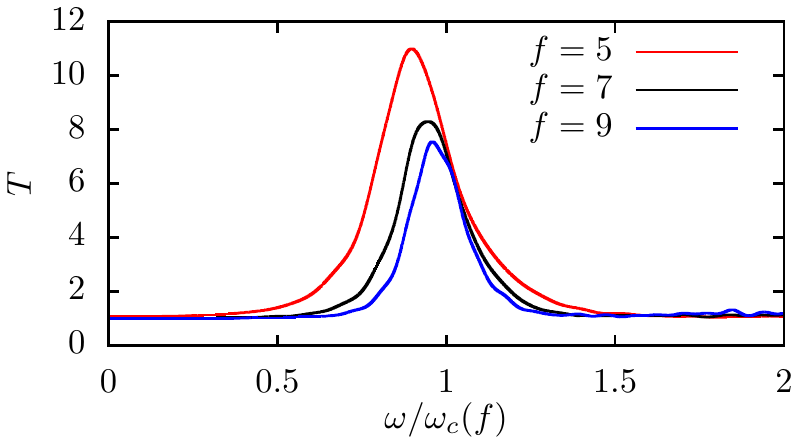}
\end{center}
\caption{Effective temperature $\Teff(\om)$ for various driving forces $f$ against frequency $\omega$ rescaled by the characteristic jump frequency $\omega_c(f)$, Eq.~\eqref{eq:om_c}.
}
\label{fig:efftemp_scaled}
\end{figure}

The frequency-dependent effective temperature $T_\mathrm{eff}(\omega)$, as defined in Eq.~\eqref{eq:teff}, is shown in Fig.~\ref{fig:efftemp}.
In equilibrium, $T_\mathrm{eff}(\omega)$ coincides with the bath temperature $T=1$ across the entire $\omega$-spectrum.
Out of equilibrium, it acquires pronounced frequency-dependent deviations, where the elevation of $\Teff(\om)$ above bath temperature can be regarded as a measure of how far the system is driven into nonequilibrium on a particular time scale $\omega^{-1}$~\cite{loi08,loi11,fodor16}.
Although the system obviously does not allow for a strict separation of time scales, salient features in $\Teff(\om)$ can still be attributed to certain aspects of the underlying dynamics via their characteristic time scales.
For instance, the mean velocity $\langle \qldot\rangle$ of the Brownian particle and the period of the potential $L$ yield a characteristic frequency
\begin{align}\label{eq:om_c}
  \omega_c(f) \equiv 2\pi \frac{\left\langle \qldot\right\rangle}{L} .
\end{align}
When the driving exceeds a critical force \mbox{$f_\mrm{c}\equiv\mrm{max}[U'(q)]$}, the Brownian particle reaches a so-called running state~\cite{riske84,reima01,lindn16}.
In this regime, the effective temperature $\Teff(\omega)$ plotted against the rescaled frequency $\omega/\omega_c(f)$, as shown in Fig.~\ref{fig:efftemp_scaled} for forces above $f_\mrm{c}=4.5$, reveals that the characteristic frequency of its main peak roughly scales with $\omega_c$.
It can hence be attributed to the impact of the periodic potential on the drifting particle.
Towards the quasi-equilibrium regime for large $f$, this peak decreases in magnitude and the effective temperature approaches the bath temperature on all time scales, \emph{i.e.}, $\mathrm{lim}_{f\to\infty} \Teff(\om)= 1$.
Noteably, for strong driving ($f=9$), $\Teff(\omega_c)$ is still many times higher than the bath temperature, despite the fact that the FDR quantities $C(\tau)$ and $R(\tau)$ in the time domain have become almost identical (\emph{c.f.} Fig.~\ref{fig:corr_resp}). When using the effective temperature as a criterion for thermalization, it is thus more appropriate to consider the FDR in the frequency domain.

In order to elicit the thermodynamical meaning of the effective temperature, we compare it to both the kinetic temperature $\Th^\mrm{kin}\equiv \mean{\qhdot^2}$ and the ``potential'' temperature $\Th^\mrm{pot} \equiv \mean{\omh^2\dqh^2}$ of the coupled oscillator.
We have used the oscillator fluctuations $\dqh$ in the definition of the latter in order to compensate for the  increasing positional bias $\mean{\qh}=-\epsilon\mean{\qldot}/\omh^2$ towards low frequencies $\omh$.
In Fig.~\ref{fig:efftemp}, both $\Th^\mrm{kin}$ (circles) and $\Th^\mrm{pot}$ (crosses) match $\Teff(\om)$ almost perfectly \emph{if} we equate the oscillator frequency $\omh$ with the frequency $\omega$ in the FDR.
Exemplarily, at $\omh=7.0$, the kinetic temperature of the oscillator is $\Th^\mrm{kin}=9.62\pm0.24$ with a corresponding effective temperature of $T_\mathrm{eff}(7.0)\simeq9.85$.
At the same time, Fig.~\ref{fig:efftemp} shows that an equipartition principle, which implies $\Th^\mrm{kin}=\Th^\mrm{pot}=\Th$, holds across the entire range of driving forces and oscillator frequencies.
The effective temperature of the driven Brownian system thus plays the same role in the thermalization of the oscillator as an equilibrium temperature would for a regular heat bath.
Conversely, the oscillator functions as a thermometer measuring the effective temperature $T_\mathrm{eff}(\omega)$ corresponding to the specific frequency $\omega=\omh$.

In the limit $\omega\to 0$, Eq.~\eqref{eq:teff} reduces to a generalized Einstein relation $\Teffbar=D_s/\mu_s$ with effective diffusivity $D_s=\int C(t) \mathrm{d}t$ and mobility $\mu_s= 2\int R(t) \mathrm{d}t$ according to the Green-Kubo relation.
For a similar model~\cite{hayas07}, this asymptotic long-term effective temperature $\Teffbar=\Teff(\omega \to 0)$ has been identified as the temperature measured by an (almost) freely moving Hamiltonian thermometer, corresponding in our model to an oscillator with $\omh\to 0$. 

\textit{Cyclic work extraction}. - Given that the oscillator thermalizes at $\Teff(\omh)$ for arbitrary eigenfrequencies $\omh$ and that equipartition holds, it effectively experiences the Brownian system as a regular heat bath albeit with variable temperature.
In order to exemplify the thermodynamical implications of this finding, we now turn the oscillator from a thermometer into a heat engine by carrying out a cyclic protocol in $\omh$.
For coupling to an equilibrium bath, such a setup has been studied earlier theoretically~\cite{schmi08} and experimentally~\cite{blick12,rossn16}.
Using in the following the notation $T_{(1,2)} \equiv T_\mathrm{eff}(\omega_{(1,2)})$, we imply that the same scheme can be applied to this nonequilibrium bath.

\begin{figure}[]
  \center
  \includegraphics{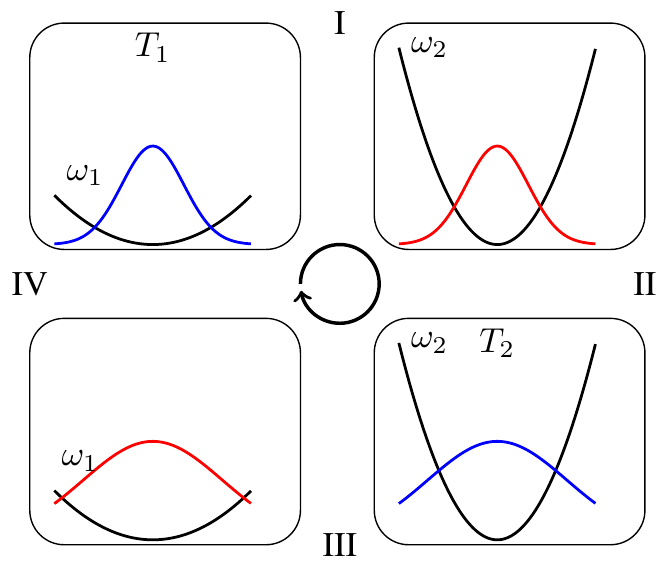}
  \caption{Scheme of the cyclic process as explained in the main text.}
  \label{fig:cycle}
\end{figure}

The cyclic process is illustrated schematically in Fig.~\ref{fig:cycle}. 
Starting out thermalized at $T_1$, in the first step (I) the oscillator is isentropically compressed by an instantaneous increase in stiffness from $\omega_1$ to $\omega_2$. 
During step (II), the oscillator thermalizes to $T_2$ at constant $\omega_2$ (isochoric).
After an isentropic and instantaneous expansion (III) from $\omega_2$ to $\omega_1$, subsequent thermalization at $T_1$ completes the cycle (IV).
This constitutes the limiting case of an Otto cycle with instantaneous compression and expansion strokes corresponding to the lower bound in efficiency, while the adiabatic (infinitely slow) cycle would yield maximum efficiency~\cite{geva92,abah14}.
Strictly separating thermalization steps from variations in stiffness facilitates the identification of work and heat. The work $\Delta W$ is readily identified as the gain in internal energy due to variations in the control parameter $\omh$ and can be calculated directly from the marginal canonical distributions
\begin{equation}
 \rho_{(1,2)}(\qh)=\sqrt{\frac{\omega_{(1,2)}^2}{2\pi T_{(1,2)}}} \,\mathrm{exp}\left[-\frac{\omega_{(1,2)}^2}{2T_{(1,2)}} \qh^2 \right] 
\end{equation}
preceding steps (I) and (III).
For a stiffness ratio \mbox{$\Omega\equiv\omega_2/\omega_1$}, the total work extracted per cycle
\begin{align}
\Delta W&=\Delta W_\mathrm{I}+ \Delta W_\mathrm{III} \\
&=\int_{-\infty}^{\infty} \!\!\!\mathrm{d}\qh \left[\rho_1(\qh)-\rho_2(\qh)\right] \left[ V_1(\qh) - V_2(\qh)\right]  \\
&=\left( 1- \Omega^2 \right) T_1/2 + \left( 1- \Omega^{-2} \right)T_2/2.
\end{align}
is given by the sum of changes in potential energy during compression and expansion steps.

Regarding the thermalization steps (II) and (IV), we point out that while conventional heat baths are by definition incapable of transferring work, a driven bath, which allows for internal currents, may also impart mechanical work to a heat engine.
However, the fact that the thermostat induces a purely thermal state in the working medium allows us to identify the change in internal energy in the second step (II) as heat
\begin{align}
\Delta Q_{\mathrm{II}}=T_2- \left(1+\Omega^2\right)T_1/2
\end{align}
flowing into the oscillator. 
Here, we have taken into account the heat ``leakage'' due to changes in mean kinetic energy, which is left unexploited by the protocol~\cite{schmi08,bauer16}.
The efficiency of the cyclic process is then properly defined as
\begin{align}
\eta\equiv\frac{\Delta W}{\Delta Q_\mathrm{II}}=1-\frac{(1+\Omega^{-2})T_2/T_1 -2}{2T_2/T_1 - (1+\Omega^2)}.
\end{align} 
It follows that the process delivers work $\Delta W>0$ only if
\begin{align}
\frac{T_2}{T_1}>\Omega^2 \quad \left(=\frac{\omega_2^2}{\omega_1^2}\right).
\end{align}
Noteably, for the driven Brownian thermostat this requires $\Teff(\om)$ to exhibit a peak as a function of frequency.
In the case at hand (\emph{c.f.} Fig.~\ref{fig:efftemp}), a sufficiently steep rise in $\Teff(\omega)$ occurs only for intermediate driving forces.
For $f=5$, we find effective temperatures $\Teff(\omega_1) \simeq 4.02$ and $\Teff(\omega_2) \simeq 7.63$ at $\omega_1=5.5$ and $\omega_2=6$, respectively. 
These parameters yield a positive efficiency of $\eta\simeq0.071$.

The cyclic process thus yields extractable work while the oscillator is in constant contact with only a single nonequilibrium thermal reservoir.
Of course, this feat is only possible due to the nonequilibrium nature of the thermostat, which has to be constantly maintained and ``paid for'' energetically by the driving. This energetic cost has been neglected here. Rather, we have treated the Brownian thermostat as given, just as one assumes ordinary heat baths as given, even though a temperature difference between them in principle also constitutes a nonequilibrium situation which has to be maintained externally. However, if one wants to account for the energetics of the driving, the dissipated heat can also be related to the FDR-based effective temperature~\cite{lippi14}.

\textit{Conclusions}. - 
In this case study, we have reinforced the relevance of an FDR-based effective temperature in the characterization of generic nonequilibrium systems by showing that it accurately predicts the thermalization properties of a coupled subsystem on any given time scale.
Probing a nonequilibrium thermostat with a harmonic oscillator, our numerical results confirm that the latter thermalizes to an effective canonical distribution and that equipartition holds.
The effective temperature of this distribution is given by the FDR of the thermostat, evaluated at the eigenfrequency of the oscillator.
With this established, we can exploit the frequency dependence of the effective temperature to realize the cyclic extraction of work from a single nonequilibrium heat bath, using the oscillator as the working substance in an Otto heat engine.
Thermodynamic considerations show that in order for the engine to yield a positive work coefficient, the effective temperature as a function of frequency has to exhibit a sufficiently steep peak.

Our results also show that, just as in ordinary thermodynamics, weak system-reservoir coupling is crucial in order to retain properties like equipartition and, where applicable, Boltzmannian statistics. We deem this relevant to the interpretation of experimental results in terms of an effective temperature, \emph{e.g.}, in the treatment of active suspensions as nonequilibrium heat baths \cite{chen07,palac10,maggi14,krish16}, where the coupling strength is not directly accessible and may in fact not be small.
On a final note, it remains to be seen whether the concept of a frequency-dependent effective temperature can be generalized to describe the thermalization of generic subsystems with a more elaborate eigenfrequency spectrum.

% ---- acknowledgments ----

\acknowledgments

We acknowledge financial support by the DFG (grant numbers SE1119/3-2 and SP1382/1-2).

% ----  bibliography ------

%

\end{document}